\documentclass{aa}
\usepackage{graphicx}	% Including figure files
\usepackage{amsmath}	% Advanced maths commands
\usepackage{amssymb}	% Extra maths symbols
\usepackage[flushleft]{threeparttable}
\usepackage[stable]{footmisc}

\newcommand{\mbh}{\ensuremath{M_{\rm BH}}}
\newcommand{\mdot}{\ensuremath{\dot{m}/\dot{m}_{\rm Edd}}}
\newcommand{\rg}{\ensuremath{r_{\rm g}}}
\newcommand{\ltransf}{\ensuremath{L_{\rm transf}/L_{\rm disc}}}
\newcommand{\fcol}{\ensuremath{f_{\rm col}}}
\newcommand{\lambdaedd}{\ensuremath{\lambda_{\rm Edd}}}
\newcommand{\ledd}{\ensuremath{L_{\rm Edd}}}
\newcommand{\lbol}{\ensuremath{L_{\rm bol}}}
\newcommand{\lxkev}{\ensuremath{L_{\rm 2keV}}}
\newcommand{\lx}{\ensuremath{L_{\rm 2-10 keV}}}
\newcommand{\luv}{\ensuremath{L_{\rm 2500\AA}}}
\newcommand{\kynsed}{{\tt KYNSED}}
\newcommand{\kxray}{\ensuremath{k_{\rm 2-10}}}

\usepackage{amsmath} % or simply amstext
\let\oldAA\AA
\renewcommand{\AA}{\text{\normalfont\oldAA}}

\usepackage[breaklinks, colorlinks, citecolor=blue, urlcolor = blue]{hyperref}
\usepackage[dvipsnames]{xcolor}
\usepackage{balance}

\usepackage{orcidlink}

\makeatletter
\renewcommand*\aa@pageof{, page \thepage{} of \pageref*{LastPage}}
\makeatother

\begin{document} 

   \title{Explaining the UV to X-ray correlation in AGN within the framework of X-ray illumination of accretion discs}
    \titlerunning{The UV to X-ray correlation in AGN}
    \authorrunning{E. Kammoun et al.}
    
\author{E. Kammoun \inst{\ref{inst1},\ref{inst3}, \ref{inst2}}\orcidlink{0000-0002-0273-218X}
\and
I. E. Papadakis \inst{\ref{inst4},\ref{inst5}}\orcidlink{0000-0001-6264-140X}
\and
M. Dov\v{c}iak \inst{\ref{inst6}}\orcidlink{0000-0003-0079-1239}
\and
E. Lusso \inst{\ref{inst7},\ref{inst3} }\orcidlink{0000-0003-0083-1157}
\and
E. Nardini \inst{\ref{inst3}}\orcidlink{0000-0001-9226-8992}
\and
G. Risaliti \inst{\ref{inst7},\ref{inst3}}\orcidlink{0000-0002-3556-977X}
} 
\institute{
Dipartimento di Matematica e Fisica, Universit\`{a} Roma Tre, via della Vasca Navale 84, I-00146 Rome, Italy \email{\href{mailto:ekammoun.astro@gmail.com}{ekammoun.astro@gmail.com}}\label{inst1}
\and
INAF -- Osservatorio Astrofisico di Arcetri, Largo Enrico Fermi 5, I-50125 Firenze, Italy\label{inst3} 
\and
Cahill Center for Astrophysics, California Institute of Technology, 1216 East California Boulevard, Pasadena, CA 91125, USA\label{inst2} 
\and
Department of Physics and Institute of Theoretical and Computational Physics, University of Crete, 71003 Heraklion, Greece \label{inst4} 
\and 
Institute of Astrophysics, FORTH, GR-71110 Heraklion, Greece\label{inst5} 
\and 
Astronomical Institute of the Czech Academy of Sciences, Bo{\v c}n{\'i} II 1401, CZ-14100 Prague, Czech Republic\label{inst6}
\and
Dipartamento di Fisica e Astronomia, Università di Firenze, via G. Sansone 1, I-50019 Sesto Fiorentino, FI, Italy\label{inst7}}
\date{Received ; accepted }
% \abstract{}{}{}{}{} 
% 5 {} token are mandatory
 
  \abstract
   {It is established that the ultraviolet (UV) and X-ray emissions in active galactic nuclei (AGN) are tightly correlated. This correlation is observed both in low- and high-redshift sources. In particular, observations of large samples of quasars revealed the presence of a non-linear correlation between UV and X-rays. The physical origin of this correlation is poorly understood.}
   {In this work, we explore this observed correlation in the framework of the X-ray illumination of the accretion disc by a central source. We have shown in previous works that this model successfully explains the continuum UV/optical time delays, variability, and the broadband spectral energy distribution in AGN.}
   {We use this model to produce $150,000$ model SEDs assuming a uniform distribution of model parameters. We compute the corresponding UV ($2500\,\AA$) and X-ray (2\,keV) monochromatic luminosities and select only the model data points that agree with the observed UV-to-X-ray correlation.}
   {Our results show that the X-ray illumination of accretion disc model can reproduce the observed correlation for a subset of model configurations with a non-uniform distribution of black hole mass (\mbh), accretion rate (\mdot), and power transferred from the accretion disc to the corona (\ltransf). In addition, our results reveal the presence of a correlation between \mbh\ and \mdot, and between \mdot\ and \ltransf, to explain the observed X-ray-UV correlation. We also present evidence based on observed luminosities supporting our findings. We finally discuss the implications of our results.}
   {}

   \keywords{Accretion, accretion discs -- Galaxies: active -- (Galaxies:) quasars: general -- (Galaxies:) quasars: supermassive black holes }

   \maketitle
%
%-------------------------------------------------------------------

\section{Introduction}
\label{sec:intro}

The correlation between the X-ray and the ultraviolet (UV) luminosities in active galactic nuclei (AGN) has been the subject of considerable interest for more than four decades. \cite{Tananbaum1979} defined the X-ray to optical luminosity ratio as $\alpha_{\rm ox} = \log \left[ l(\nu_{\rm X-ray})/l(\nu_{\rm opt})\right]/\log\left(\nu_{\rm X-ray}/\nu_{\rm opt} \right)$, where $l(\nu)$ is the monochromatic luminosity (in ${\rm erg\,s^{-1}\,Hz^{-1}}$) at rest-frame frequency $\nu$. Since then, this ratio has been extensively used to explore the X-ray--UV connection in quasars \citep[e.g.][]{Avni1982, Avni1986, Kriss1985, Wilkes1994, Starteva2005, Lusso2010, Young2010, Lusso2016}. These works examined the correlation between the monochromatic $2\,\rm keV$ and $2500\,\AA$ luminosities. This reduces the X-ray to optical ratio to $\alpha_{\rm ox} =  0.3838 \log \left( \lxkev / \luv \right)$.  In all cases, it is generally found that $\alpha_{\rm ox}$ is anti-correlated with the UV luminosity \citep[see e.g.][]{Lusso12}. This anti-correlation is interpreted as the by-product of the observed non-linear correlation between \lxkev\ and \luv\ of the form $\log \lxkev = \gamma \log \luv + \beta$, with $\gamma \sim 0.5-0.7$ \citep[see e.g.][]{Vignali2003, Starteva2005, Steffen2006, Young2010, Lusso2016}. This implies that optically bright AGN generally emit fewer X-rays (per unit UV luminosity) than optically faint sources.

It is commonly thought that AGN UV/optical photons originate from the accretion disc \citep{Shields1978}, and the X-ray continuum emission is the result of inverse Compton scattering of these disc photons in a hot `corona' \citep[e.g.][]{Galeev1979, Haardt1994}. Thus, understanding the origin of the relationship between the UV and X-ray emission is important to test the energy generation mechanism in AGN, and to explore the influence of different physical parameters to the observed properties of AGN. We note that some works investigated alternative wavelengths as proxies of the X-ray/UV correlation \citep[see e.g.][]{Signorini2023, Jin2024}, finding that monochromatic luminosities at 2\,keV and 2500\,\AA\ are reasonable proxies of the X-ray corona and disc emission in AGN. In addition, this relationship can be used to derive bolometric corrections for AGN and estimate luminosity and accretion rate with the goal of studying the evolution of accretion onto supermassive black holes (SMBHs) through cosmic time \citep[see e.g.][]{Young2010} .

\cite{Dovciak22} presented \kynsed\footnote{\url{https://projects.asu.cas.cz/dovciak/kynsed/-/tree/main}}, a new model to estimate the broadband optical/UV/X-ray spectral energy distribution (SED) of an AGN when X-rays illuminate the disc. \kynsed\ assumes a \cite{Novikov73} accretion disc illuminated by a point-like X-ray corona. \cite{Dovciak22} and \cite{Kammoun2024} demonstrated that this model can explain the time-averaged and the variable optical/UV/X-ray SED of NGC\,5548 as well as its timing properties, during its long monitoring campaign in 2014 \citep{Derosa15}. In addition, the X-ray illumination model can explain the observed UV/optical continuum time lags \citep[]{Kammoun19lag, Kammoun21a, Kammoun21b, Kammoun23, Langis2024} as well as the UV/optical power spectra \citep{Panagiotou20, Panagiotou22, Papoutsis2024} in local AGN, and the disc-size problem in micro-lensed quasars \citep{Papadakis2022}.

In this paper, we study the correlation between the X-ray (2\,keV) and UV (2500\,\AA) luminosity observed in many samples of quasars \citep[e.g.][]{Lusso2020}. Our goal is to investigate whether X-ray disc illumination can explain the observed X-ray/UV correlation in AGN. In Sect.\,\ref{sec:simulation} we present the model and the predicted theoretical $\lxkev-\luv$ correlation when we consider a broad range of physical parameter values, all of them following a uniform distribution. In Sects.\,\ref{sec:observation} and \ref{sec:results} we compare the model predictions with the observed X-ray/UV correlation in AGN, and we find that X-ray disc illumination can explain the observed correlation but only for a smaller sub-sample of the physical parameters, with specific correlations between BH mass and accretion rate, as well as between the power of the X--ray source and accretion rate and BH mass. These correlations, which are necessary for the model to explain the observed X-ray/UV relation in AGN, imply certain links between the main physical parameters of AGN (like BH mass and accretion rate) and measurable quantities such as the X-ray, optical, and bolometric luminosity. In Sect.\,\ref{sec:predictions} we show that the model predictions are in agreement with what is observed in AGN. We summarise our results and we discuss their implications in Sect.\,\ref{sec:discussion}.

%%%%%%%%%%%%%%%%%%%%%%%%%%%%%%%%%%%%%%%%%%%%%%%%%%%%%%%%%%% FIG 1.
\begin{figure*}
    \centering
    \includegraphics[width=0.95\linewidth]{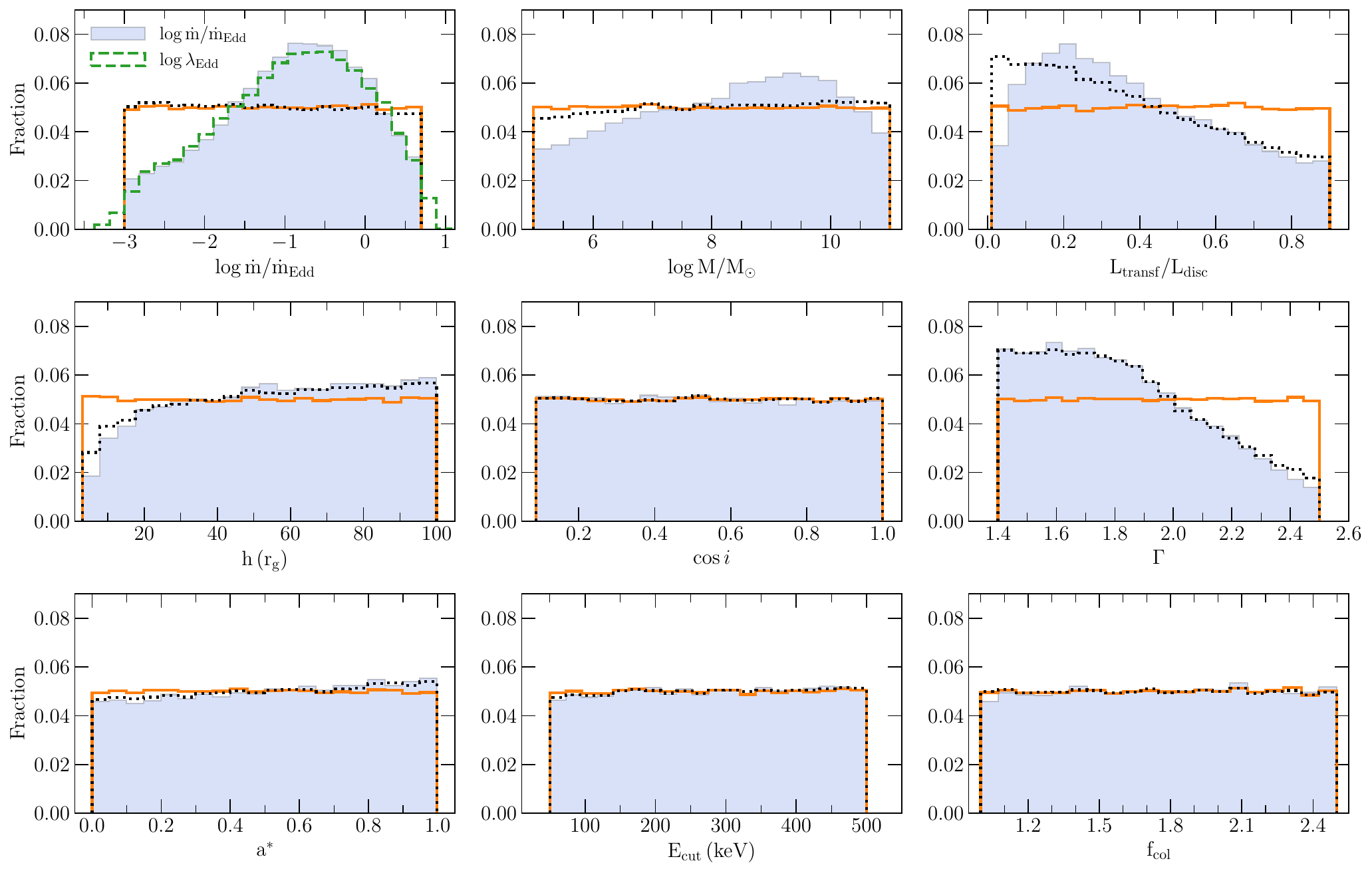}
\caption{Distribution of each of the model parameters of the full set of model SEDs (orange solid lines), the subset of SEDs where the size of the corona is smaller than the height of the source (black dotted lines), and of the selected subset of SEDs that agree with the observed $\lxkev-\luv$ correlation (blue filled  histograms). This subsample consists of 28\% ($42,000$ SEDs)of the full set of model SEDs. The green dashed histogram in the upper left panel corresponds to the distribution of $\log \lambdaedd$ of the selected subset of model SEDs.}
\label{fig:histograms}
\end{figure*}
%
%%%%%%%%%%%%%%%%%%%%%%%%%%%%%%%%%%%%%%%%%%%%%%%%%%%%%%%%%%%%%%%%%%%%%%%%%% Fig. 2
\begin{figure}
    \centering
    \includegraphics[width=0.99\linewidth]{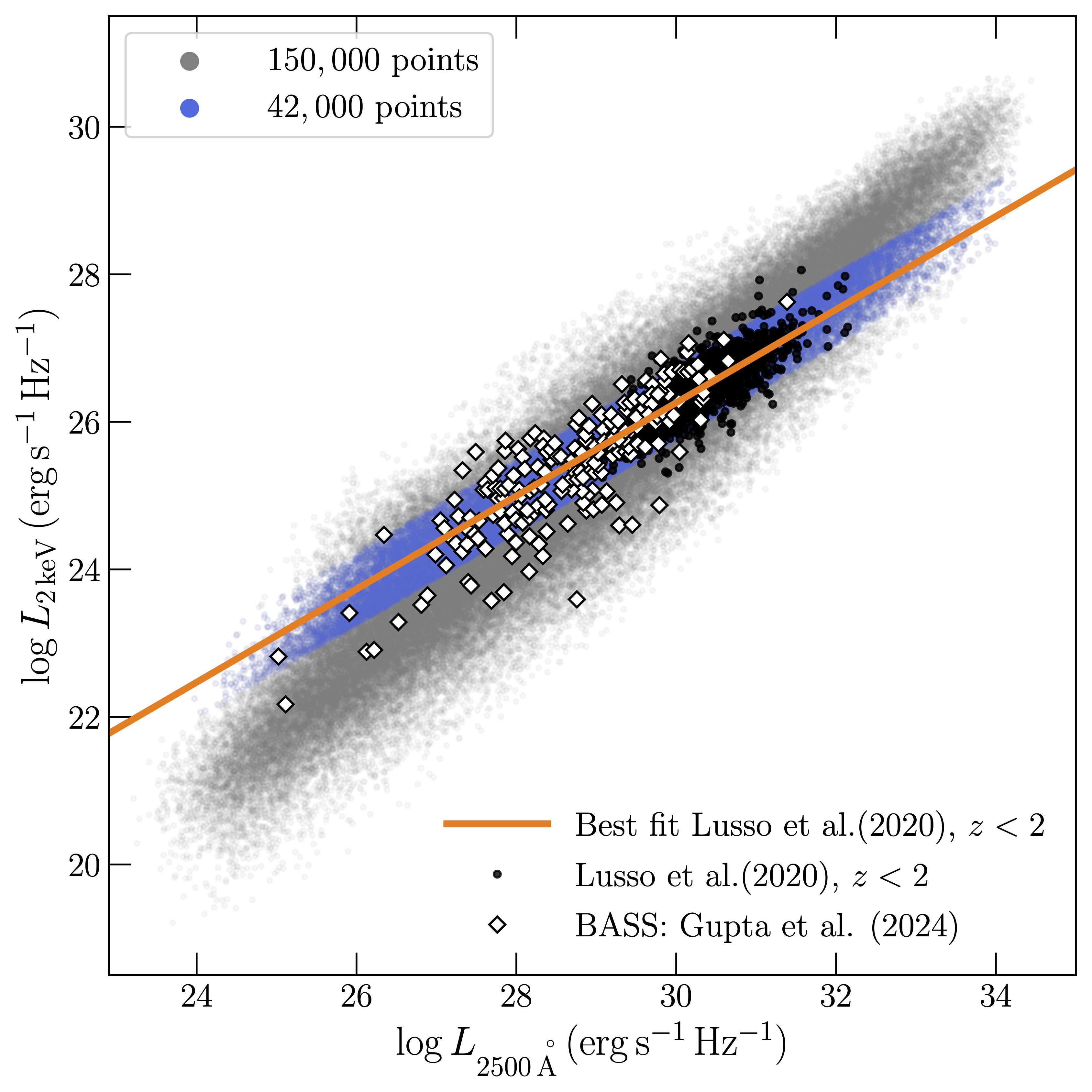}
\caption{X-ray luminosity versus UV luminosity for the full set of model SEDs where the corona radius is physical (grey points; see Sect.\,\ref{sec:simulation} for more details). The black points correspond to the data obtained from \citet{Lusso2020} for quasars at redshift below 2. The orange line shows the best-fit straight line fitted to the black points. The blue points correspond to the model SEDs that are consistent with the best-fit line within $2\sigma$. White diamonds indicate the data from the BASS sample \citep{Gupta2024}}
\label{fig:simulations}
\end{figure}

%%%%%%%%%%%%%%%%%%%%%%%%%%%%%%%%%%%%%%%%%%%%%%%%%%%%%%%%%%%%%%%%%%%%%% FIG. 3
\begin{figure*}
\centering
\includegraphics[width=\linewidth]{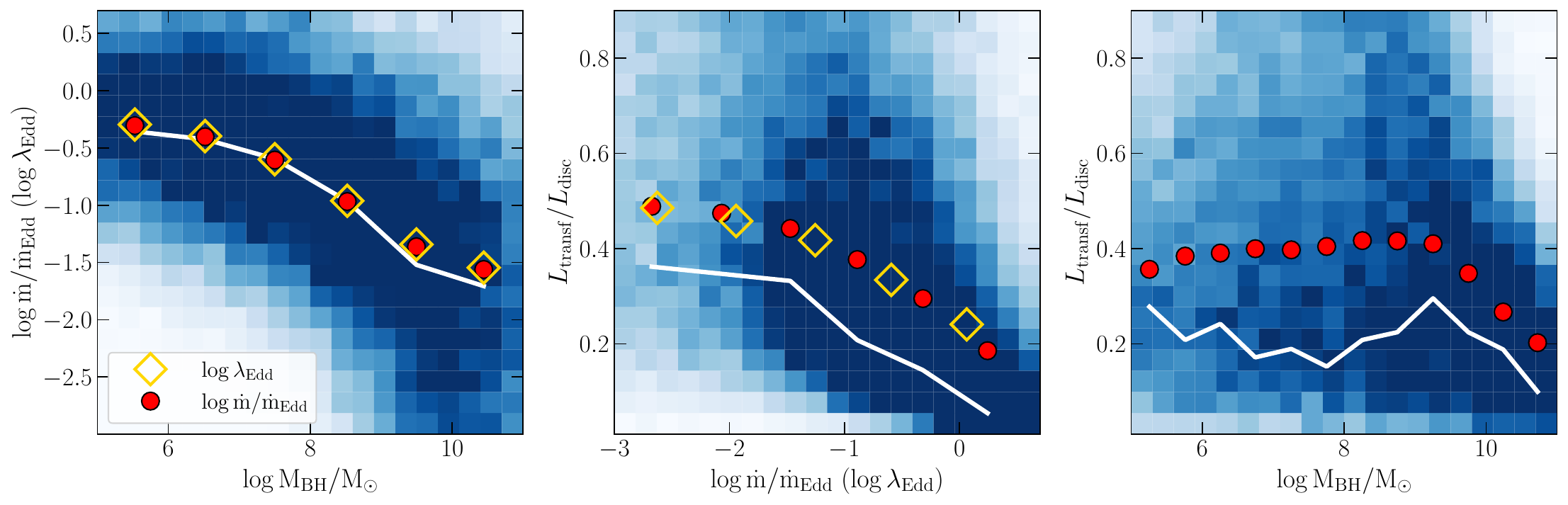}

\caption{Correlation between $\log \mdot$ and $\log \mbh$, \ltransf\, and $\log \mdot$, and between \ltransf\, and $\log\mbh$ (left, middle and right panels respectively) for the subset of model SEDs that are consistent with the observed $\lxkev-\luv$ correlation. Blue squares correspond to the 2D histograms. The red circles correspond to the median parameters in narrow bins along the $x$-axis. The empty diamonds show the median $\log \lambdaedd$, computed in the same bins. The white solid lines show the mode of the distribution in each of the bins along the $x$-axis. }
\label{fig:M-mdot-Ltransf}
\end{figure*}

%%%%%%%%%%%%%%%%%%%%%%%%%%%%%%%%%%%%%%%%%%%%%%%%%%%%%%%%%%%%%%%%%%%%%%%%%%%%%%%%%%%%%%%%%%%%%%%%%%%%%%%%%%%%%%%%%%%%%%%%%%

%%%%%%%%%%%%%%%%%%%%%%%%%%%%%%%%%%%%%%%%%%%%%%%%%%%%%%%%%%%%%%%%%%%%%%%%%%%%%%%
%%%%%%%%%%%%%%%%%% Section 2.
\section{Theoretical \texorpdfstring{$\lxkev-\luv$}{Lx-Luv} correlation}
\label{sec:simulation}

%%%%%%%%%%%%%%%%%%%%%%%%%%%%%%%%%%%%%%%%%%%%%%%%%%%%%%%%%%%%%%%%%%%%%
%%%%%%%%%%%%%%%%%%%%%%  TABLE 1.
\begin{table}
\centering
\caption{Range of model parameters used to compute the model SEDs.}
\label{tab:parameters}
\begin{tabular}{lr} % four columns, alignment for each
\hline \hline
Parameter        & Range	                \\ \hline
$\log \mbh/M_\odot$	    &	    [$5, 11$]    	\\ [0.1cm]
$\log \mdot$		        & [$-3,0.7$ ]	        \\ [0.1cm]
$a^\ast$		        & [$0, 0.998$ ]	        \\ [0.1cm]
\ltransf 	        & [$0.01,0.9$] 	           \\ [0.1cm]
$h\, (\rg)$	        & [$3,100$ ]        \\ [0.1cm] 
$\cos i$	        & [$0.088,0.999$]       \\ [0.1cm]            
$\Gamma$	        & [$1.4,2.5$]                    \\ [0.1cm]  
$E_{\rm cut}\, (\rm keV)$	        & [$50,500$]	  \\[0.1cm]
\fcol	        & [$1,2.5$]               \\ [0.1cm]  
\hline
\end{tabular}
\end{table}
%%%%%%%%%%%%%%%%%%%%%%%%%%%%%%%%%%%%%%%%%%%%%%%%%%%%%%%%%%%%%

\kynsed\ provides the theoretical, broadband SED, as emitted by the accretion disc and the X-ray corona, in the case when the corona illuminates the disc in the lamp-post geometry. It takes into account all relativistic effects as well as the feedback between the disc and the corona \citep{Dovciak22}.  Within this framework, the disc reprocesses the incident X-rays and re-emit part of this radiation in the form of the so-called reflection spectrum, in X-rays. The remaining radiation is absorbed by the disc, leading to an increase in its temperature. This will result in a UV/optical light radiation in excess of the standard disc emission, that will be emitted by the disc after a certain delay. 

The model can provide SEDs both in the case when the X--ray corona is powered by the accretion process itself, and when it is powered externally, by an unknown source not directly associated with the accretion power. In the former case, the accretion power dissipated in the disc below a given transition radius, $r_{\rm tr}=R_{\rm tr}/R_g$\footnote{$R_g=G$\mbh$/c^2$ is the gravitational radius of a BH with a mass of \mbh.}, is transferred to the X-ray source (by a yet unspecified physical mechanism). In this way, a direct link between the accretion disc and the X--ray source is set. The fraction of the accretion power dissipated within $r_{\rm tr}$ is referred to as \ltransf. In the following, we assume that the X-ray corona is powered by the accretion disc. The model allows also the use of a colour-correction factor, \fcol, for the disc emission. This factor (whose value is unknown for AGN) accounts for the disc photons scattering off the free electrons present in the upper layers of the accretion disc, which can lead to deviations from the blackbody emission.

The main physical parameters of \kynsed\, are: the black hole (BH) mass, \mbh, the accretion rate in units of the Eddington accretion rate, \mdot, the BH spin, $a^{\ast}$, \ltransf, the height of the corona, $h$, the inclination angle (parameterized as $\cos i$), the X-ray spectrum photon index, $\Gamma$, a high-energy cutoff energy, $E_{\rm cut}$, and the colour correction factor, \fcol. In order to study the theoretical relation between the X-ray and UV luminosity, we used \kynsed\ to simulate $150,000$ model SEDs\footnote{We produce the model SEDs using {\tt PyXSPEC}, the Python interface of {\tt XSPEC} \cite{Arnaud96}.} assuming a uniform distribution for all parameters within the ranges stated in Table\,\ref{tab:parameters}. This is the simplest assumption we can make for the distribution of the parameters. The parameter distributions are shown as orange histograms in Fig.\,\ref{fig:histograms}. 

In all cases, we considered a range of values wider than what has been observed in large AGN samples, just in case current surveys have missed AGN with parameters either smaller and/or larger than the observational limits that have been established so far. Consequently, regarding the accretion rate, most AGN accrete at the $\sim 0.01$ up to 1 of their Eddington limit \citep[at least in the local universe; see e.g.][]{Koss2022cat}, and we considered a larger range, between 0.001 and 5 times the Eddington limit. Similarly, most of the nearby AGN host BHs with a mass $\sim 10^6-10^9\, M_{\odot}$ \citep{Koss2022cat}, although masses up to $\sim 10^{10}\,M_{\odot}$ have been measured in higher redshift objects \citep[see e.g.][]{Duras20}. We therefore considered a slightly larger sample of BH masses between $10^5-10^{11}\, M_{\odot}$. The chosen range of $\Gamma$ (between $1.4-2.5$) is driven by the fact that most of the sources in \cite{Lusso2020}, who analysed a large sample of selected quasars from the Sloan Digital Sky Survey (SDSS), have $\Gamma \lesssim 2.5$, while most of the sources in the \textit{Swift}/BAT AGN Spectroscopic Survey (BASS) sample \citep{Ricci2017, Gupta2024} have $\Gamma \gtrsim 1.4$. As for the high energy cut-off in the X-ray spectra of AGN, observations indicate that $100\,{\rm keV} \lesssim E_{\rm cut}\lesssim 300\,\rm keV$ \citep[see e.g.][]{Ricci2018}. We investigate a slightly larger range, between 50 and 500\,keV. 

BH spin, corona height, \ltransf, inclination and \fcol\, are parameters that cannot be easily constrained by observations. In fact, one of the objectives of this project is to investigate their distribution for the X-ray disc illumination model to explain the observed X-ray/UV correlation in AGN. For that reason, we consider a uniform distribution of these parameters over all of the possible values (note that in the case of \ltransf\, we cannot consider values larger than unity, as we assume that the X-ray corona is powered by the accretion process, while \fcol\ cannot be much larger than 2.5 \citep[see e.g.][]{Ross92}. 

As discussed in Sect.\,2.2 of \cite{Dovciak22}, \texttt{KYNSED} computes the corona radius ($R_{\rm c}$) a posteriori for a given set of parameters, assuming conservation of energy and photon number. All the assumptions and the method of calculating $R_{\rm c}$ are detailed in \cite{Dovciak2016} and \cite{Dovciak22}. Consequently, a set of parameters can be considered physical only if the corona can fit outside of the event horizon of the BH. In other terms, $R_{\rm c}$ should be smaller than the difference between the corona height and the radius of the event horizon. Thus, we selected the SEDs where this condition is fulfilled, which consist of $\sim 70\%$ of the original sample (a total of $103,000$ SEDs out of $150,000$). The parameter distribution of this subset is shown as dotted histogram in Fig.\,\ref{fig:histograms}. This figure shows that the distribution of \ltransf, $h$, and $\Gamma$ deviates significantly from a uniform distribution. The spin distribution shows a slight deviation. The distribution of all of the other parameters remains consistent with being uniform. The deviations in the distributions we see are due to the fact that at low heights if \ltransf\ were to be large, this would imply a high luminosity which requires a large radius. This explains the deficit in low height and high \ltransf. Similarly, for large values of \ltransf, if the corona had soft X-ray spectrum this would imply a large radius which explains the deficit in the distribution at high values of $\Gamma$.

For each of the model SEDs, where the corona radius is physical, we compute the monochromatic luminosity at 2\,keV and $2500$\,\AA\ ($\lxkev$ and $\luv$, respectively), the total observed luminosity ($\lbol$, by integrating the SED between $\sim 20\,\rm \mu m$ and 1000\,keV), the $2-10$\,keV luminosity ($\lx$),  and $\lambdaedd = \lbol / \ledd$, which we refer to as the `Eddington ratio'. We note that the \lambdaedd\ differs from \mdot\ depending on the disc inclination angle. This connection is further explored in Appendix\,\ref{sec:lambdaEddmdot}. Grey dots in Fig.\,\ref{fig:simulations} show $\log \lxkev$ as a function of the corresponding $\log \luv$ for the $150,000$ model SEDS. Clearly, under the hypothesis of a uniform distribution of the physical parameters, X-ray disc illumination predicts that \lxkev\ should increase proportionally with \luv, and the slope of the correlation between $\log \lxkev$ and $\log \luv$ should be $\sim 1$.

%%%%%%%%%%%%%%%%%%%%%%%%%%%%%%%%%%%%%%%%%%%%%%%%%%%%%%%%%%%%%%%%%%%%%%%%%%%%%%%
%%%%%%%%%%%%%%%%%% Section 3.
\section{The observed \texorpdfstring{$\lxkev-\luv$}{Lx-Luv} correlation}
\label{sec:observation}

We accept the results of \citet{Lusso2020} as being representative of the X-ray/UV correlation in AGN. \citet{Lusso2020} presented a sample of quasars up to $z \sim 7$ where the intrinsic \luv\ and \lxkev\ could both be measured. The sources were chosen to show low levels of dust reddening, host galaxy contamination, X-ray absorption, or Eddington bias effects, as discussed in their Sect.\,7. In order to avoid any inaccuracy in estimating the intrinsic luminosity due to the assumed cosmology, we consider only the sources with redshift below 2 from the clean sample of \citet{Lusso2020}. 

We use the standard ordinary least square method ($\rm OLS(Y|X)$) from \citet{Isobe1990}, to fit a straight line ($y_{\rm model} = \beta + \gamma x$) to these sources in the $\log \lxkev-\log\luv$ plane. We normalize the fit at $\log\luv = 30.5$. This results in a slope $\gamma = 0.631 \pm 0.012$, and a normalization of $\beta = 26.576 \pm 0.007$. The scatter\footnote{We calculate the scatter as: $\sigma = \sqrt{\sum_{i=1}^{i=N} \frac{(y_i - y_{\rm model, i})^2}{N}} $.} of the data points around the best-fit line is $\sigma = 0.24\,\rm dex$. The data from \citet{Lusso2020} and the best-fit line are shown in Fig.\,\ref{fig:simulations} as black dots and orange solid line, respectively. 

White diamonds in the same figure show the (\lxkev, \luv) data for Type-1 sources from the BASS sample \citep{Gupta2024}. Clearly, the X-ray/UV data for the unobscured AGN in the nearby Universe are in agreement with the data for the distant quasars. They appear to be also consistent with the best-fit orange line, and their scatter around this line is similar to the scatter of the \citet{Lusso2020} points. The \citet{Lusso2020} and the BASS sample data points in Fig.\,\ref{fig:simulations} indicate that the X--ray/UV correlation in AGN, as defined by the orange line in this figure, extends over (at least) five orders of magnitude in UV luminosity, from $\sim 27$ to 32 in $\log \luv$. 

Interestingly, the observed data points appear to be a subset of the model data points. We note that there was no reason, a priori, for such an agreement between the model and the data. However, it is clear that the slope of the best-fit line to observed $\log \lxkev-\log \luv$ data deviates from the slope of 1 that is expected in the case of the X-ray disc reflection model, if the parameters were uniformly distributed as indicated by the orange histograms in Fig.\,\ref{fig:histograms}.

In the following section, we investigate which subset of the parameter space can actually produce the observed $\lxkev-\luv$ correlation, within the framework of our model. To do so, we selected the model data points in Fig.\,\ref{fig:simulations} which are consistent within $2\sigma$ with the best-fit line to the data. These model points, indicated by the blue points in the same figure, consist of 28\% of the full set of simulated SEDs ($42,000$ SEDs out of the original $150,000$ simulated ones). We consider all the points, over the full range of the X-ray and UV luminosity resulting from the model SEDs. This is larger than the range of the X-ray and UV luminosity sampled by the BASS and the \cite{Lusso2020} observations, but we keep the full range in case the observations so far have not revealed the full population of AGN. We note that choosing points with any scatter smaller than $3\sigma$ around the best fit would not affect the conclusions of our results (see Sect.\,\ref{sec:test_sigma} for more details.)

In the following section, we consider the model parameters which correspond to the X-ray and UV luminosity pairs indicated by the blue points in Fig.\,\ref{fig:simulations}. Our objective is to investigate their distribution and potential correlations which are necessary to explain the observed $\lxkev-\luv$ correlation. 

%%%%%%%%%%%%%%%%%%%%%%%%%%%%%%%%%%%%%%%%%%%%%%%%%%%%%%%%%%%%%%%%%%%%%%%%%%%%%%%%%%%%%%%%%%%%%
%%%%%%%%%%%%%%%%%%%%%%%%%%%%%%%%%%%%%%%%%%%%%%%%%%%%%%%%%%%%%%% SECTION 3.

\section{The model parameters that can explain the observed X-ray/UV correlation}
\label{sec:results}

The blue filled histograms in Fig.\,\ref{fig:histograms} show the distribution of the parameters for the selected subset of model SEDs which are consistent with the observed  $\lxkev-\luv$ correlation. The parameter distributions of this subset agree with the original distributions (dotted histograms in Fig.\,\ref{fig:histograms}) except for $\log \mbh $, $\log \mdot$, and \ltransf. The new distribution of $\log \mbh/M_\odot $ shows a broad peak at $\sim 8.5-10$. The abrupt cuts at 5 and 11 are due to the limits we assumed for the original set of model SEDs. The distribution of \ltransf\ deviates from the original distribution at low values (below $\sim 0.4$), where a peak can be seen at 0.2. The distribution of $\log \mdot$ is the one which deviates the most from the original uniform distribution. It shows a clear peak at accretion rates $\log \mdot$ between $\sim -1$ and $-0.5$ (as before with the BH mass distribution, the abrupt cuts at $-3$ and $0.7$ are the limits we originally set for the model SEDs). The green dashed-line in the top-left panel in Fig.\,\ref{fig:histograms} shows the resulting distribution for $\log \lambdaedd$. It is very similar to the $\log \mdot$ distribution, but it shows a more gradual decline towards very large and small values (i.e. beyond $\log\lambdaedd$ of 0.5 and $-3$, respectively). 
 
Fig.\,\ref{fig:M-mdot-Ltransf} shows the 2D histogram of the accretion rate vs BH mass, and of \ltransf\ vs accretion rate and BH mass (left, middle and right panel, respectively). The colour scheme is chosen so that the intensity of the bins increases with the number of model points in each bin. In order to better illustrate the correlation between the various parameters, we considered the distribution of $\log \mdot$ (left panel) and \ltransf\ (middle and right panels) in narrow bins along the $x$-axis. The filled circles indicate the median value in each bin. The open diamonds show the same but considering $\log \lambdaedd$ instead of $\log \mdot$. Given the fact that the distributions in each bins may not be symmetric, we also show the mode of the distributions as white solid lines. Clear correlations can be seen between all three parameters. The left panel shows that \mdot\ decreases with increasing BH mass. As for \ltransf, the middle panel shows that it remains relatively constant at $\sim 0.4-0.5$, on average, for $\log \mdot \lesssim -1$, and then it decreases with increasing accretion rates. On the other hand, \ltransf\ does not appear to correlate with BH mass, up to $\log\mbh \sim 9$. At higher BH masses, \ltransf\ decreases with increasing BH mass. The mode of the distributions perfectly agrees with the median in the left panel. In the middle and the right panel, the trends seen in the mode and the median are similar in shape, but with a difference in the amplitude, which is due to the asymmetry of the distributions.

Our main result so far is the following: the observed correlation between the X-ray and UV luminosity is a generic feature of the X-ray illumination of the accretion disc model. If we assume that this correlation extends over the full luminosity range, as indicated by the blue points in Fig.\,\ref{fig:simulations}, then it can be reproduced as long as: (a) the distributions of \mbh, \mdot, and \ltransf\ follow the blue histograms in Fig.\,\ref{fig:histograms}, and (b) \mdot\ correlates with \mbh\ and \ltransf\ correlates with \mdot\ and \mbh\ as shown in Fig.\,\ref{fig:M-mdot-Ltransf}.

%%%%%%%%%%%%%%%%%%%%%%%%%%%%%%%%%%%%%%%%%%%%%%%%%%%%%%%%%%%%%
%%%%%%%%%%%%% Fig. 3
\begin{figure}
    \includegraphics[width=\linewidth]{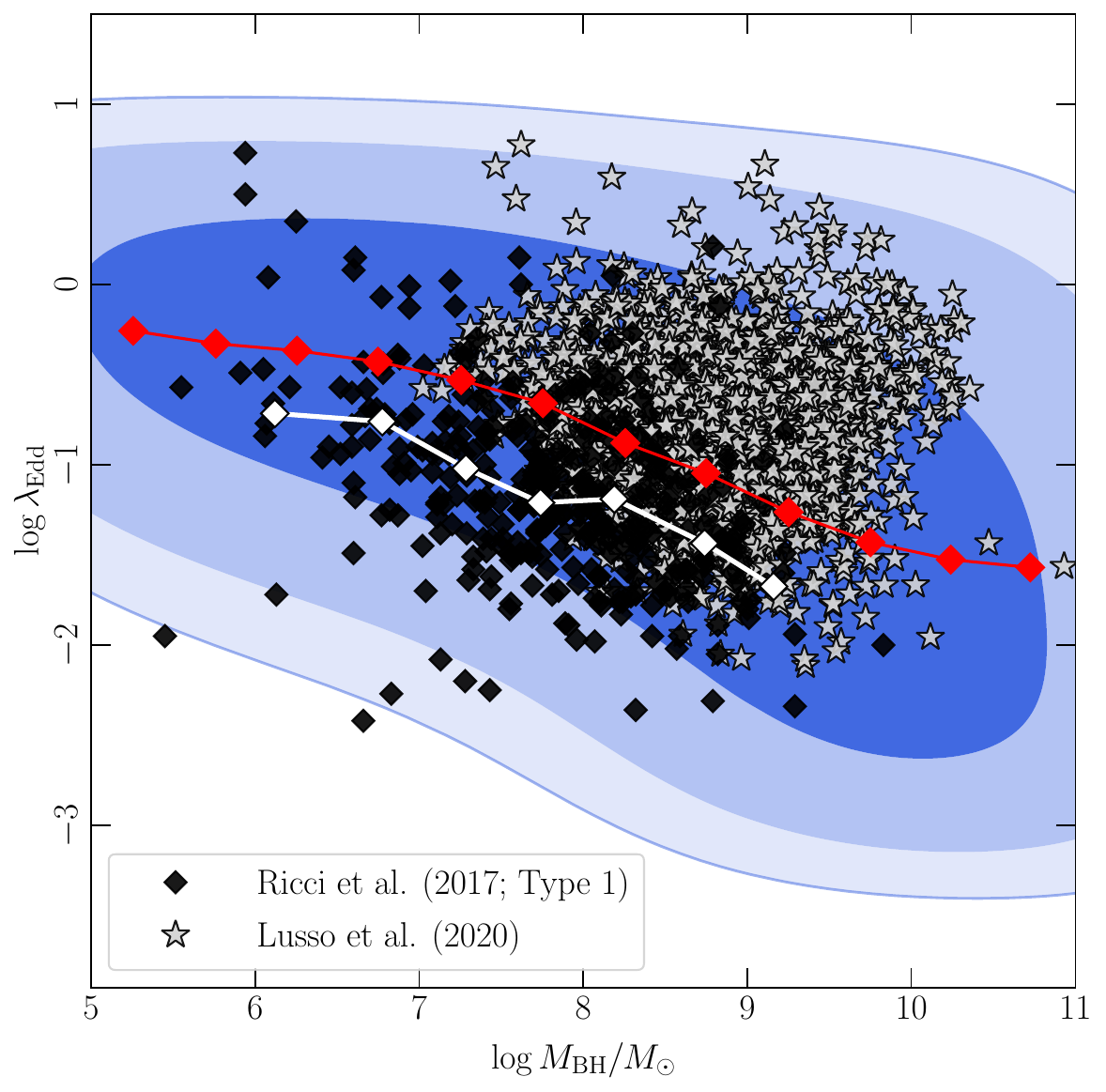}
    \caption{Plot of the Eddington ratio versus black hole mass for the subset of model SEDs that agree with the $\lxkev-\luv$ correlation. Darker to lighter blue shaded contours correspond to the 68\%, 95\%, and 99\% of the model data points. We also show the observed data from \citet[grey stars]{Lusso2020}, from the 
    BASS sample (only Type-1 sources; black diamonds). Red and white connected diamonds indicate the median data from the subsample of SEDs that agree with the observed $\lxkev-\luv$ correlation and the ones from the BASS sample, respectively.}
    
    \label{fig:contour}
\end{figure}
%%%%%%%%%%%%%%%%%%%%%%%%%%%%%%%%%%%%%%%%%%%%%%%%%

%%%%%%%%%%%%%%%%%%%%%%%%%%%%%%%%%%%%%. SECTION 4
\section{Observational validity tests of the X-ray disc illumination model predictions}
\label{sec:predictions}

The agreement between the model and the observed $\lxkev-\luv$ correlation does not prove that the X-ray disc illumination model is indeed the correct one for AGN. However, our results imply that the X-ray disc illumination model can explain the observed correlation only if certain correlations exist between the accretion rate, BH mass, and \ltransf. These correlations are specific to the X-ray illumination model. We do not expect AGN to necessarily follow these correlations, if other models are responsible for the observed $\lxkev-\luv$ correlation in these objects.

For this reason, we present below the results from the comparison between the model specific predictions (shown in Fig.\,\ref{fig:M-mdot-Ltransf}) with observations. In particular, we consider the physical parameters for the model SEDs for which \lxkev\ and \luv\ agree with the observed correlation in AGN. As we already mentioned, there is no reason, a priori, why the correlation that the model predicts between  physical parameters like \mdot\ and \mbh, will be in agreement with the corresponding observed correlations, if a different physical mechanism is responsible for  the UV/X-ray correlation in AGN. Thus, the observational tests we present in the following are independent ways to investigate the validity of the X-ray disc illumination model.

%%%%%%%%%%%%%%%%%%%%%%%%%%. 4.1
\subsection{Observational evidence for the \texorpdfstring{\lambdaedd}\, vs \texorpdfstring{\mbh}\, relation in AGN}
\label{sec:lambdaEddMass}

The Eddington ratio is one of the key AGN properties as it helps understand the accretion process and is a crucial parameter for AGN feedback. In practice, one can compute \lambdaedd, which is a good estimate of \mdot\, when the observed emission is not strongly affected by absorption and if the inclination angle is not larger than $\sim 40-50^\circ$. In Fig.\,\ref{fig:contour}, we show the distribution of \lambdaedd\ as a function of \mbh\ in the selected subset of model SEDs that agree with the observed $\lxkev-\luv$. Instead of showing the data points themselves, in this figure we show contours which encompass 68\%, 95\%, and 99\% of our model data.

The grey stars in this figure show the $(\log \mbh, \log \lambdaedd)$ data points for the quasars in \cite{Lusso2020} with $z < 2$. For these sources, we consider the corresponding \mbh\ and \lambdaedd\ from \cite{Wu2022} who provided measurements of various quasar properties from the SDSS data release 16. Clearly the \cite{Lusso2020} quasars are in full agreement with the model predictions, as the majority of the sources lies within the 68 per cent contour of the model data. 

We can compare our results with data from other recent AGN surveys as well. For example, black diamonds in Fig.\,\ref{fig:contour} show the $(\log \mbh, \log \lambdaedd)$ data for Type-1 AGN from the \textit{Swift}/BAT AGN Spectroscopic Survey \citep[BASS;][]{Koss2017,Koss2022cat, Koss2022}. This is a hard X--ray selected sample, which may not be affected by strong selection effects, and is representative of the AGN population in the local Universe. We show data for the Type-1 objects only, to be able to compare them with results from other samples which consist of Type-1 AGN, mainly. In any case, Type-1 sources are preferable, as our modeling does not account for absorption. Fig.\,\ref{fig:contour} shows that the majority of the BASS data are consistent with the 68\%\ contour of our model prediction. Furthermore, white diamonds in this figure indicate the mean $\log \lambdaedd $ vs $\log\mbh$ relation in the BASS data, which shows the same shape as the relation predicted by our model (red diamonds in the same figure). The difference in amplitude between the white and red diamonds is due to the selection effects that could be present in the BASS sample, especially the fact that BASS is less sensitive to sources with high Eddington ratios, as they tend to have softer X-ray spectra.

%%%%%%%%%%%%%. FIG 5.
\begin{figure}
\centering
\includegraphics[width=0.99\linewidth]{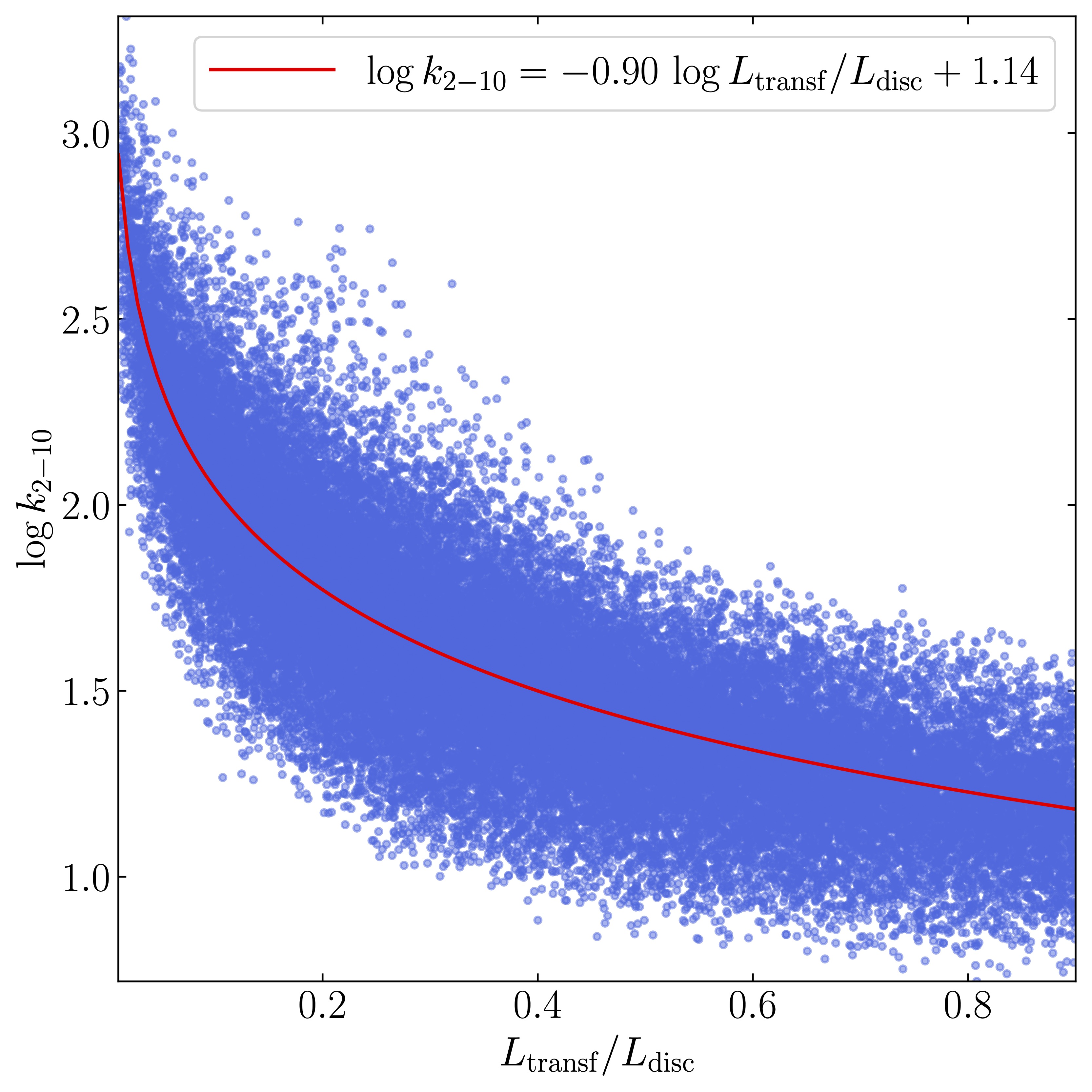}
\caption{X-ray bolometric correction as a function of \ltransf\ for the subset of model SEDs that agree with the observed \lxkev-\luv\ correlation. The red line shows the best-fit line to this correlation.}
\label{fig:kx_ltransf}
\end{figure}
%%%%%%%%%%%%%%%%%%%%%%%%%%%%%%%%%%%%%%%%%%%%%%%%%%%%%%%%%%%%%%%%%%%%%%%%%%%%%%%%%%%%%%%%%%%%%%%%%%%%%%%%%%%%%%%%%%%%%%%%%%

The results in this section indicate that a relation between the accretion rate and the BH mass exists for the quasars in the \cite{Lusso2020} sample that we used to define the $\lxkev-\luv$ relation, and for other samples as well. In all cases, this relation is entirely consistent with the predictions of the disc X-ray illuminated model. We investigate below additional observational evidence which shows that AGN are also consistent with the other model requirement to explain the observed $\lxkev-\luv$ relation, i.e. the \ltransf\ vs \mdot\ and \mbh\ correlations shown in the middle and right panels in Fig.\,\ref{fig:M-mdot-Ltransf}.

%%%%%%%%%%%%%%%%%%%%%%%%%%%%%%%%%%%%%%%%%%%%%%%%%%%%%%%%%%%%%%%%%%%%%%%%%%%%%%%%%%%%%%%%%%%%%%%%%%%%%%%%%%
%%%%%%%%%%%%%. FIG 5.
\begin{figure}[ht!]
\centering
\includegraphics[width=0.9\linewidth]{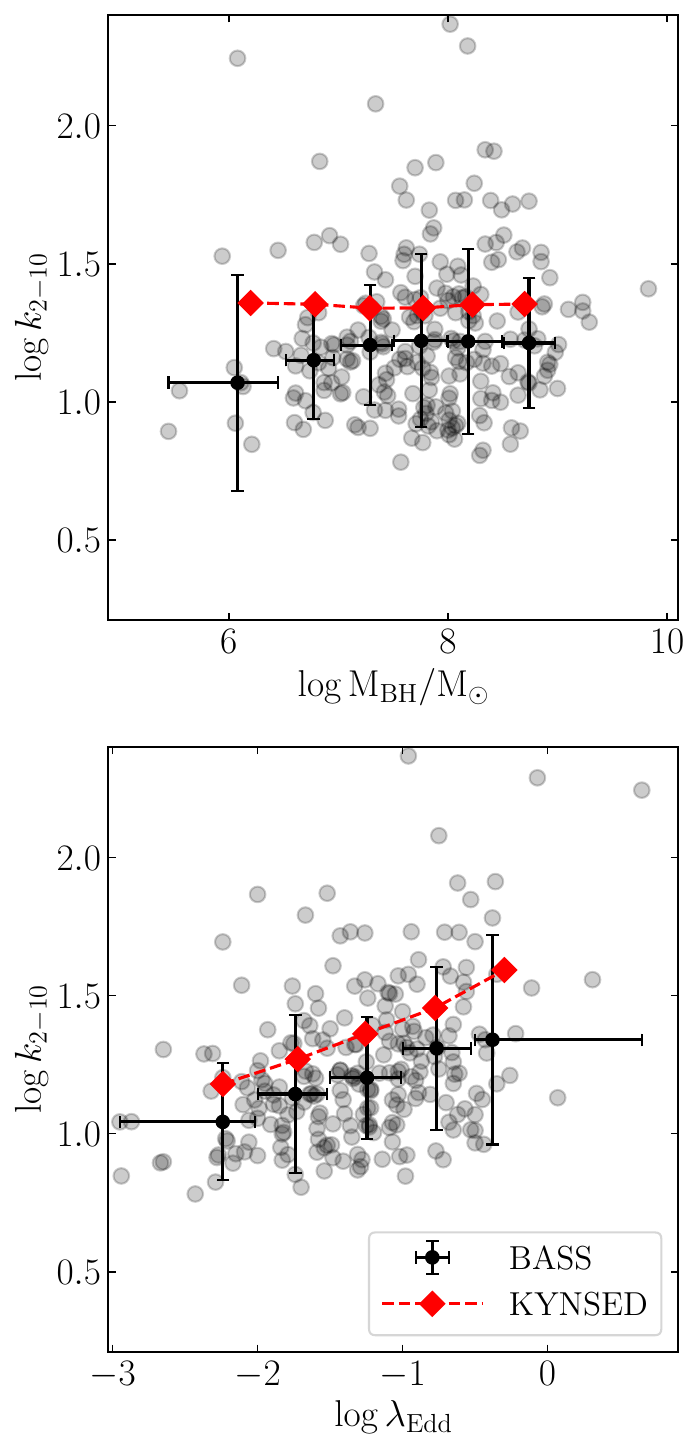}
\caption{X-ray bolometric correction as a function of \mbh\ (top) and \lambdaedd\ (bottom). The grey circles correspond to the data points from BASS \citep{Gupta2024}. The black circles show the same data but binned along the $x$-axis. We also show \kxray\ as a function of \mbh\ and \lambdaedd\ for the model SEDs that agree with the observed \lx-\luv\ correlation obtained by matching the distributions of \mbh\ and \lambdaedd\ to the ones of the BASS sample (red diamonds).}
\label{fig:kx_M_mdot}
\end{figure}
%%%%%%%%%%%%%%%%%%%%%%%%%%%%%%%%%%%%%%%%%%%%%%%%%%%%%%%%%%%%%%%%%%%%%%%%%%%%%%%%%%%%%%%%%%%%%%%%%%%%%%%%%%%%%%%%%%%%%%%%%%

\subsection{Observational evidence for the \texorpdfstring{\ltransf-\mdot}\, and \texorpdfstring{\ltransf-\mbh}\, relations in AGN}
\label{sec:bolcorr}

As we already mentioned, it is not easy to measure \ltransf\ in AGN from observations. However, it is rather trivial to determine the $2-10$\,keV luminosity in these objects, and consequently the X-ray bolometric correction ($\kxray = \lbol/\lx$). These two quantities are expected to be negatively correlated as \ltransf\ represents the ratio of the X-ray luminosity to the accretion luminosity. Fig.\,\ref{fig:kx_ltransf} shows that indeed this negative correlation is seen in the model SEDs that agree with the observed \lxkev-\luv\ correlation. We fit this correlation in the log-space and found a slope of $-0.9$. We note that the same correlation can be seen if we consider the full set of model SEDs. Thus, we use in this section \kxray\ as a proxy for \ltransf\ to test the dependence on mass and accretion rate.

In Fig.\,\ref{fig:kx_M_mdot}, we show \kxray\ as a function of \mbh and \lambdaedd, for the Type-1 AGN observed in the BASS sample \citep{Gupta2024}. As mentioned earlier, we expect the dependence of \kxray\ on \lambdaedd\ and \mbh\ to be consistent with the ones shown in the middle and right panel of Fig.\,\ref{fig:M-mdot-Ltransf}, respectively. Thus, for the \mbh\ range probed by BASS ($\log \mbh \sim 6-9$), we do not expect to see any change in \kxray. However, we expect \kxray\ to increase with \lambdaedd\ (\kxray\ being inversely proportional to \ltransf). The black circles in this figure correspond to the binned BASS data along the $x$-axes, and show similar trends to the one expected, albeit with a relatively large scatter. The red diamonds in this figure show the same quantities from the model SEDs that agree with the observed \lx-\luv\ correlation, considering a similar binning. These points are obtained by selecting the model SEDs that match the observed \mbh\ and \lambdaedd\ distributions from the BASS sample. This figure shows a clear agreement between our model values and the ones observed in the BASS sample. In Appendix\,\ref{sec:observedlum}, we show how the $2-10$\,keV luminosity depends on the BH mass, accretion rate, and \ltransf, which is another way of quantifying the effects discussed in this section.

%%%%%%%%%%%%%%%%%%%%%%%%%%%%%%%%%%%%%%%%%%%%%%%%%%%%%%%%%%%%%%%%%%%%%%%%%%
%%%%%%%%%%%%%%%%%%%%%%%% SECTION 4.

\section{Discussion and Conclusions}
\label{sec:discussion}

In summary, we used the {\tt KYNSED} model, which assumes X-ray illumination of accretion disc, to estimate $150,000$ pairs of AGN X-ray and UV luminosity data points for a wide range of physical parameters, assuming a uniform distribution of a reasonable range of values. We found that only a sub-sample of the model parameters can predict an X-ray and UV luminosity in agreement with the observed $\lxkev-\luv$ correlation  presented by \cite{Lusso2020}. Our results indicate that the observed $\lxkev-\luv$ correlation is a generic feature of X-ray disc illumination as long as two conditions are fulfilled: (a) \mbh, \mdot, and \ltransf\ follow the distributions shown in Fig.\,\ref{fig:histograms} as blue histograms, and (b) \mdot\ correlates with \mbh\ and \ltransf\ correlates with \mdot\ and \mbh\ as shown in Fig.\,\ref{fig:M-mdot-Ltransf}. In addition, we found evidence that the correlations predicted by our model are in full agreement with AGN observations, thus supporting, further, the X-ray illuminated disc scenario in AGN. We note that our findings depend on the coronal geometry, which is assumed to be a lamp post in \texttt{KYNSED}. \cite{Dovciak22} showed that considering a 3D spherical corona would have a minimal effect compared to a point source corona. A different coronal geometry (e.g., a radially extended corona) would alter the illumination pattern of the disc, which will consequently modify the disc response and the thermally reprocessed radiation. Investigating different coronal geometries goes beyond the scope of this work. However, we stress that any geometry must be able to explain all of the observed features in AGN, i.e., the time lag behavior as a function of wavelength, the observed variability and correlations between various wavelengths, and the spectral properties.

\subsection{On the predicted model correlations}
\label{sec:test_sigma}
%%%%%%%%%%%%%%%%%%%%%%%%%%%%%%%%%%%%%%%%%%%%%%%%%%%%%%%%%%%%%%%%%%%%%%%%%%%%%%%%%%%%%%%%%%%%%%%%%%%%%%%%%%%%%%%%%%%%%%%%%%
\begin{figure*}
\centering
\includegraphics[width=0.85\linewidth]{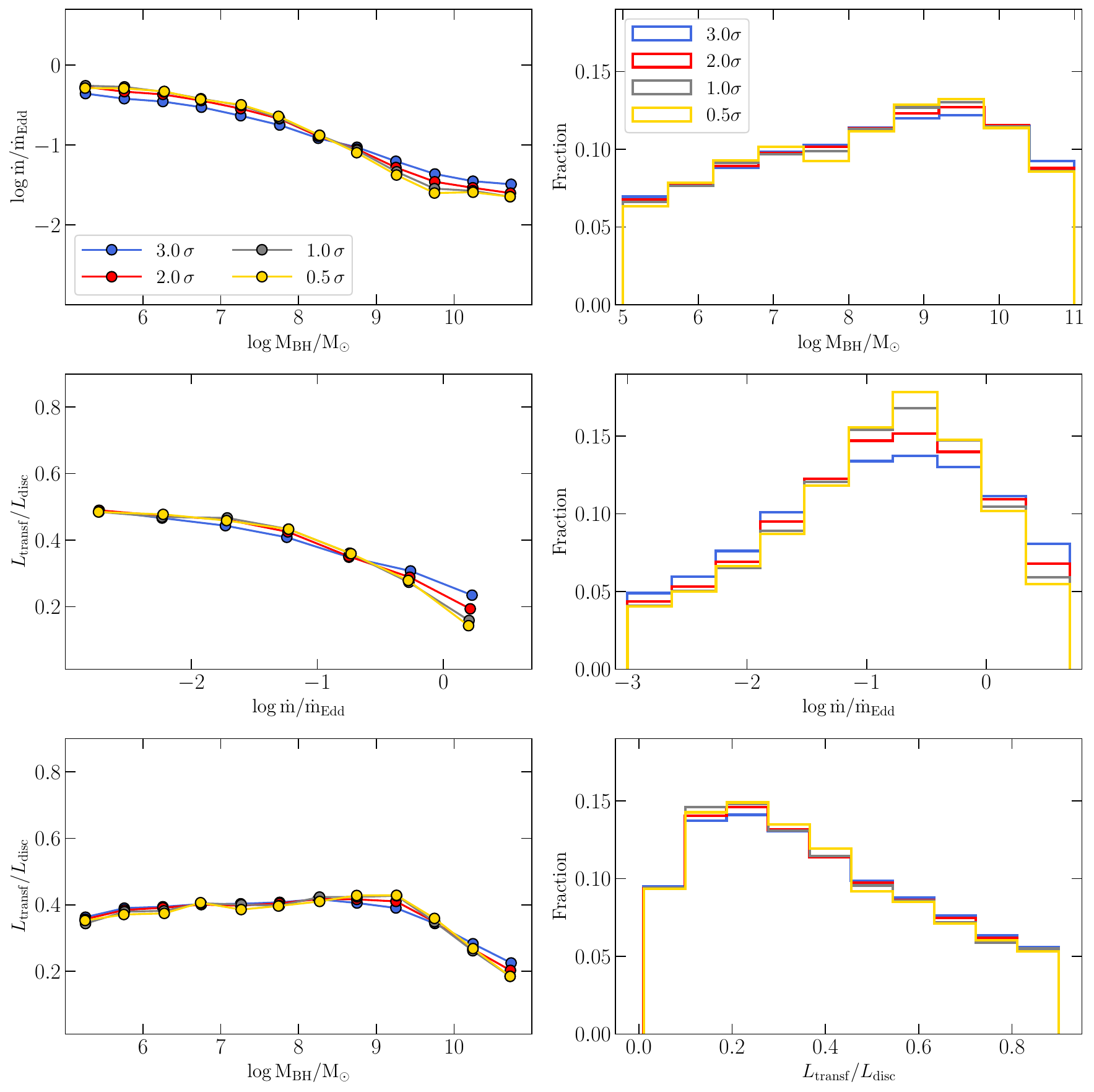}

\caption{Left: Average correlations between $\log \mdot$, $\log \mbh$, and \ltransf, similar to Fig.\,\ref{fig:M-mdot-Ltransf} considering different values of the scatter around the best-fit \lxkev-\luv\ correlation ranging between $0.5\sigma$ and $3\sigma$. Right: Distribution of $\log\mbh$, $\log \mdot$, and \ltransf\ (top to bottom) for the selected subset of SEDs that agree with the observed $\lxkev-\luv$ correlation assuming a scatter around the best-fit between $0.5\sigma$ and $3\sigma$.}
\label{fig:correlation_sigma}
\end{figure*}
%%%%%%%%%%%%%%%%%%%%%%%%%%%%%%%%%%%%%%%%%%%%%%%%%%%%%%%%%%%%%%%%%%%%%%%%%%%%%%%%%%%%%%%%%%%%%%%%%%%%%%%%%%%%%%%%%%%%%%%%%%

The correlations between \mbh, \mdot, and \ltransf\ presented in Fig.\,\ref{fig:M-mdot-Ltransf} are one of the main predictions of this work. These correlations as well as the distributions of the model parameters (blue histograms in Fig.\,\ref{fig:histograms}) depend on a) the choice of the range of each of the parameters in Table\,\ref{tab:parameters}, and b) the choice of the $2\sigma$ scatter around the best fit (blue points in Fig.\,\ref{fig:simulations}). While the true range of each of the model parameter values cannot be known a priori, we believe that our choice is a good representation of the range of parameters observed in AGN. Regarding the choice of the $2\sigma$ scatter around the best-fit $\lxkev-\luv$ correlation, we tested how different values of the scatter could affect the predicted correlations. 

In the left panel of Fig.\,\ref{fig:correlation_sigma}, we show that the average correlation between the three parameters is not affected by the choice of the scatter. We considered scatter between $0.5\sigma$ and $3\sigma$ around the best-fit $\lxkev-\luv$ correlation shown in Fig.\,\ref{fig:simulations}. Obviously the correlations become less important as the scatter increases. However, Fig.\,\ref{fig:correlation_sigma} shows that the predicted correlations do not change much for a scatter below $ 3\sigma$. This figure shows that our choice of the $2\sigma$ scatter does not affect the conclusions of this work. Furthermore, this also shows that our conclusions still hold even if the intrinsic scatter around the $\lxkev-\luv$ correlation is very small. In fact, some works argue that the dispersion in the $\log \lxkev - \log \luv$ correlation could be much smaller than the value of $\sim 0.24\,\rm dex$, found by \cite{Lusso2020} \citep[see e.g.][]{Sacchi2022, Signorini2024}. In the right panel of Fig.\,\ref{fig:correlation_sigma}, we show the 1D distribution of \mbh, \mdot, and \ltransf, for the model SEDs that agree with the observed $\lxkev-\luv$ correlation assuming a scatter around of the best fit below $3\sigma$. This shows that the distributions of the parameters are not highly affected by the choice of the scatter as long as it is below $3\sigma$. 

\subsection{The independent, additional tests}
The quasars from the \cite{Lusso2020} sample, which we used to define the observed $\lxkev-\luv$ correlation, follow the $\lambdaedd-\mbh$ correlation (and consequently the $\mdot-\mbh$ correlation) predicted by our model as shown in Fig.\,\ref{fig:contour}. The agreement between our model predictions and observed data also applies to all of the samples we considered in Sect.\,\ref{sec:predictions}. In particular, the agreement between our model predictions and the BASS sample is a strong validation of our model, since this sample is supposedly representative of the total population of AGN in the local Universe. Furthermore, we show in Sect.\,\ref{sec:bolcorr} that our model predicts a specific dependence of \kxray\ on \mbh\ and \lambdaedd, that agrees with the trends observed in the BASS sample. As we mentioned in this section, the correlation is a proxy for the \ltransf-\mbh\ and \ltransf-\mdot\ correlations. Observations of AGN samples result in correlation that are in complete agreement with the correlation predicted by our model, thus supporting, further, the X-ray illuminated disc scenario in AGN.

\subsection{Implications of the \mbh, \mdot, and \ltransf\ correlations}

The correlation between \mbh\ and \mdot\ can be explained in the context of BH growth via accretion. Various works showed that the accretion episode of a SMBH starts with a high accretion rate (in units of Eddington) that decreases as the BH grows in mass and consumes its reservoir \cite[see e.g.][]{Granato2001, Granato2004, Merloni2004, Lapi2006}. Such models have been proposed to explain the observational properties of different types of quasars and the co-evolution of SMBHs and their host galaxies \citep[e.g.][]{Kawakatu2003,Kawakatu2007}. It is worth noting that \cite{Lusso2020} showed that the $\lxkev-\luv$ correlation is valid in all of the redshift ranges probed in their study (see their Fig.\,8 and Appendix\,B). This means that the correlations we predict between \mbh, \mdot, and \ltransf\ should be valid in all redshift ranges as well, and consequently all evolutionary stages of the sources (if quasars would just grow via accretion). We note that our model does not consider the presence of AGN feedback, which could also affect the SMBH evolution and the observed properties of AGN. 

The correlation between \ltransf\ and \mdot\ must be indicative of the unknown physical mechanism that powers the X-ray corona. Our model predicts that the fraction of power transferred to the X-ray corona compared to the disc power decreases at high accretion rates. This behavior is consistent with what is expected in the case of a magnetic reconnection-heated corona scenario \citep[see for example Fig.\,3 in][]{Liu2016magrec}. This is also supported by results from studying X-ray bolometeric corrections in sample of AGN, showing that the X-ray to bolometric luminosity fraction decreases at high luminosity \citep[see e.g.][]{Lusso12, Duras20}.

In conclusion, we show in this paper that X-ray illumination of accretion discs explains the observed correlation between the X-rays and UV in AGN. Our results also shed light on the physical properties and the evolution of AGN, as we predict very specific correlations between the BH mass, accretion rate, and \ltransf that are needed to produce the observed $\lxkev-\luv$ correlation in AGN. All of these model predictions agree with the observations of large sample of AGN. 

\subsection*{Data availability}
The data that support the findings of this study are available from the corresponding author, EK, upon reasonable request.

\begin{acknowledgements}
We thank the anonymous referee for their constructive feedback. EK and IEP acknowledge support from the European Union’s Horizon 2020 Programme under the AHEAD2020 project (grant agreement n. 871158). Software packages used in this study include {\tt XSPEC/PyXSPEC} \citep{Arnaud96}, {\tt NumPy} \citep{Numpy}, {\tt SciPy} \citep{Scipy}, {\tt Pandas} \citep{PandasPaper}, {\tt Matplotlib} \citep{Matplotlib}, and {\tt GetDist} \citep{Lewis19}.
\end{acknowledgements}


\bibliographystyle{aa} % style aa.bst
\bibliography{references} % your references Yourfile.bib

%%%%%%%%%%%%%%%%%%%% APPENDICES %%%%%%%%%%%%%%%%%%
\begin{appendix}
\onecolumn
\section{The connection between the Eddington ratio and the accretion rate}
\label{sec:lambdaEddmdot}

\begin{figure*}[h!]
\centering
\includegraphics[width=0.9\linewidth]{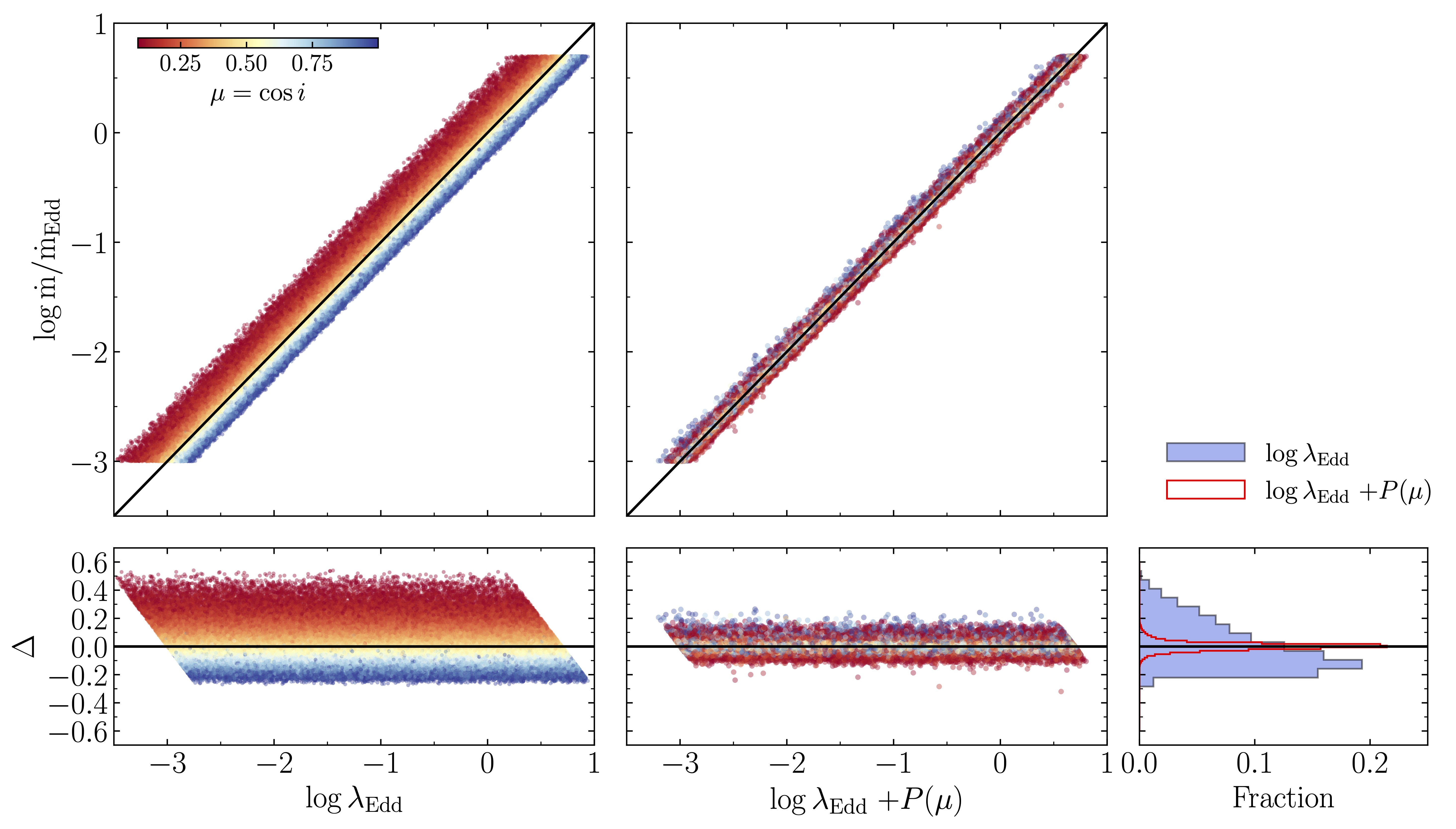}
\caption{ Left: The accretion rate ($\log \mdot$) as a function of the Eddington ratio ($\log \lambdaedd$). The solid line show the identity line. The bottom panel shows the difference between $\log \mdot$\ and the identity line. Middle: Same but replacing the $x$-axis with the sum of $\log \lambdaedd$ and a polynomial function of the inclination angle to account for the anisotropy in the emission (see text for more details). The color maps corresponds\ to $\mu = \cos i$. The bottom right panel shows the distribution of the differences between the $y$-axis and the $x$-axis in the left and middle panels (blue and red histograms, respectively)}
\label{fig:mdot_lambda_cosi}
\end{figure*}

In this section, we explore the connection between \mdot\ and \lambdaedd. Within {\tt KYNSED}, \mdot\ is accretion rate normalized to the Eddington accretion rate. The accretion power is computed self-consistently within the code by considering the radiative efficiency ($\eta$), which is a function of the BH spin ( $ L = \dot{M} \eta(a^\ast) c^2$). If the emission were to be isotropic, \mdot\ should be equal to \lambdaedd. However, since we see a dependence on the inclination angle as shown in the left panel Fig.\,\ref{fig:mdot_lambda_cosi}. We see a non-symmetric and relatively large scatter compared to the 1:1 correlation. To quantify the dependence on inclination, we divided the points into bins of $\cos i$ with a width of 0.01. Then we fitted $\log \mdot$ vs $\log \lambdaedd$ in each bin with a straight line. As expected, the best-fit slopes are consistent with 1 for all the bins, with an average value of 0.997. However, the normalizations vary as a function of inclination. We fitted the normalization as a function of $\cos i$ with a phenomenological polynomial of degree three:

\begin{equation}
  P(\mu) = -0.673 \mu^3 + 1.703 \mu^2  - 1.739\mu + 0.503, 
\end{equation}

\noindent where $\mu = \cos i$. When we add the effect of inclination, the scatter becomes symmetric and reduces to 0.03\,dex. Thus, based on our model, for a given inclination angle, the relation between \mdot\ and \lambdaedd\ can be described as:

\begin{equation}
 \log \mdot = \log \lambdaedd + P(\mu)\, \pm 0.03. 
\end{equation}

\noindent Thus, for a given source, knowing the value of \lambdaedd\ based on observations, the above equations could be used to estimate the accretion rate assuming a certain inclination angle.

%%%%%%%%%%%%%. FIG A.1
%%%%%%%%%%%%%%%%%%%%%%%%%%%%%%%%%%%%%%%%%%%%%%%%%%%%%%%%%%%%%%%%%%%%%%%%%%%%%%%%%%%%%%%%%%%%%%%%%%%%%%%%%%%%%%%%%%%%%%%%%%
%%%

\section{Observational evidence for the \texorpdfstring{\lx-\mbh-\mdot}\, relations in AGN}
\label{sec:observedlum}

%%%%%%%%%%%%%. FIG B.1
\begin{figure*}[ht!]
\centering
\includegraphics[width=0.9\linewidth]{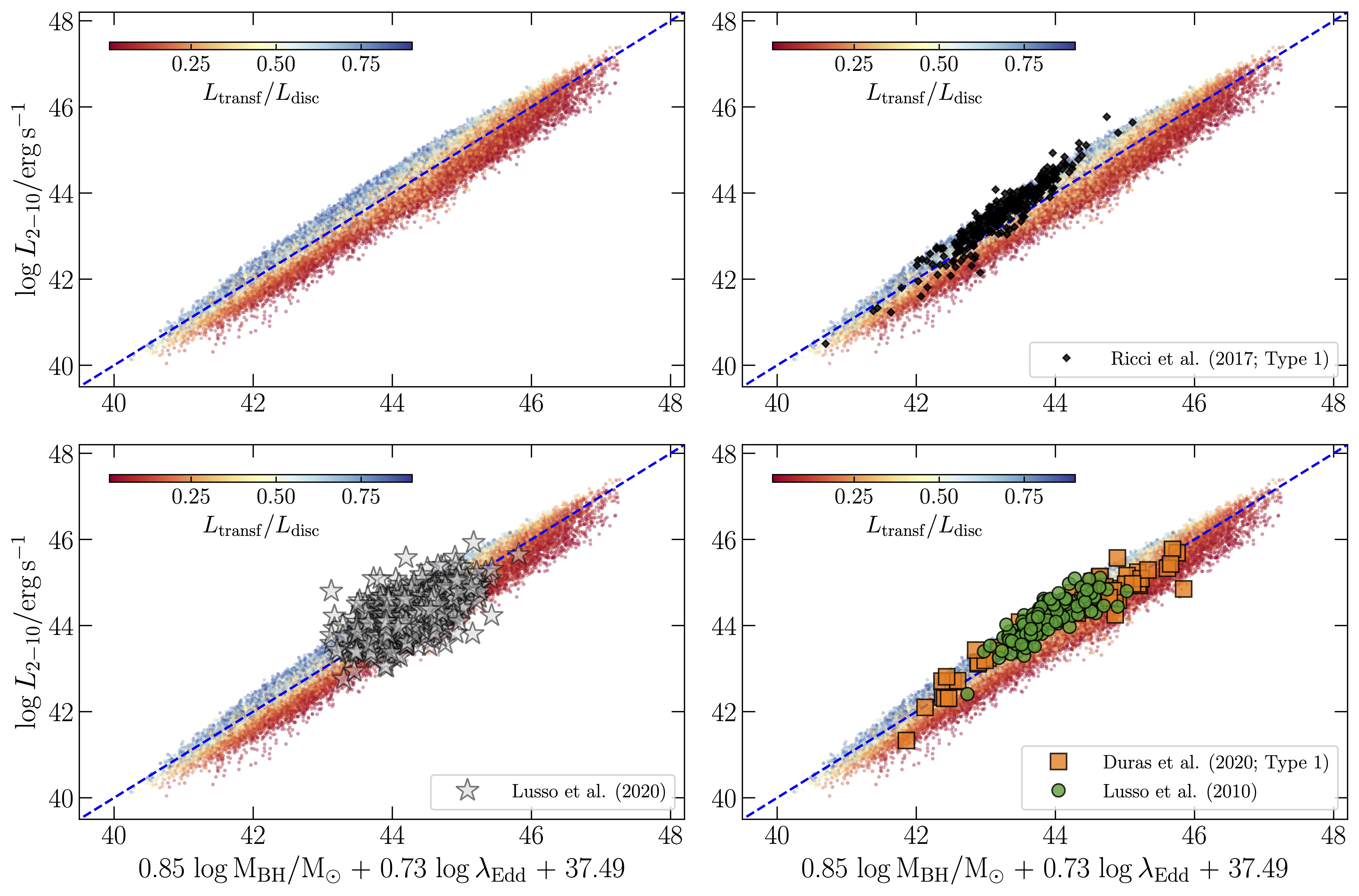}
\caption{X-ray luminosity as a function of the BH mass and Eddington ratio for the subset of model SEDs that agree with the observed \lxkev-\luv\ correlation. Color maps show the distribution in \ltransf. Each of the panels show the data points from the observed samples we consider in this work.}
\label{fig:lx_M-mdot}
\end{figure*}
%%%%%%%%%%%%%%%%%%%%%%%%%%%%%%%%%%%%%%%%%%%%%%%%%%%%%%%%%%%%%%%%%%%%%%%%%%%%%%%%%%%%%%%%%%%%%%%%%%%%%%%%%%%%%%%%%%%%%%%%%%

In this section, we investigate whether there is a correlation between \lx\ and \mdot/\mbh\ in the model SEDs which are consistent with the observed $\lxkev-\luv$ correlation in AGN. To this end, we fitted $\log \lx$ as a linear combination of $\log \mbh$ and $\log \lambdaedd$ using the Scipy function {\tt curve\_fit}\footnote{\url{https://docs.scipy.org/doc/scipy/reference/generated/scipy.optimize.curve_fit.html}}. The best-fit correlation obtained can be written as follows:

\begin{equation}\label{eq:lx-m-mdot}
    \log \lx = 0.85 \log \mbh + 0.73 \log \lambdaedd + 37.49. 
\end{equation}

\noindent Fig.\,\ref{fig:lx_M-mdot} shows $\log \lx$ plotted as a function of the best-fit linear relation defined by this correlation. Clearly, the X-ray illuminated disc model predicts a very strong correlation between \lx\ and \mbh/\lambdaedd\ in AGN, with a $1\sigma$ scatter of 0.26\,dex around the best fit. The color map in this figure shows the distribution of \ltransf. Roughly speaking, AGN with larger \ltransf\ should be above the dashed line in this plot, while AGN with smaller \ltransf\ should be below.

We also show in Fig.\,\ref{fig:lx_M-mdot} the correlation between the observed X-ray luminosity versus the correlation with \mbh\ and \lambdaedd\ we found in Eq.\eqref{eq:lx-m-mdot} using available data from the different samples we considered in the previous section. Clearly, the observed correlation agrees extremely well with the model predictions. The top right panel shows the data from the Type\,1 sources of the BASS sample by \cite{Ricci2017} and revised by \cite{Gupta2024}. Interestingly, the majority of these sources fall above the blue line, in agreement with high \ltransf\ values. This is also the case for the data from \cite{Lusso2010} and \cite{Duras20}, shown in the bottom right panel. As these three samples are X-ray selected, it is thus most likely that they are biased towards high values of \ltransf, where the X-ray power is larger. In fact, X-ray selected samples tend to show an X-ray bolometric correction of the order of $\sim 10-30$, which indeed corresponds to high values of \ltransf\ from Fig.\,\ref{fig:kx_ltransf}. The bottom left panel of this figure shows the data points obtained by \cite{Lusso2020}. These data points fill a larger scatter around the best fit compared to the other samples.

We note that, in general, a correlation between between the X-ray luminosity and \mbh\ as well as \mdot\ should be expected for any model that predicts the observed $\lxkev-\luv$ correlation in AGN. After all,  the 2\,keV luminosity is tightly correlated with \lx, while \luv\ depends on \mbh\ and \mdot. However, in the X-ray illuminated disc model, the source of heating for the inner accretion disc is the absorbed X-rays, which are not exactly equal to the accretion power that is transferred to the corona. Therefore, according to the model, \luv\ should also depend on \ltransf\ itself. Consequently, the model correlation plotted in Fig.\,\ref{fig:lx_M-mdot} is very specific to our model, and is representative of the X-ray illuminated disc but only for the model parameters which can explain the X-ray/UV correlation in AGN.

\end{appendix}

%-----------------------------------------------------------------------------
\label{LastPage}
\end{document}